\documentclass{jfm}
\usepackage{amsmath}  
\usepackage{amssymb}  
\usepackage{mathtools}
\usepackage{float}

\usepackage{placeins}
\usepackage{tikz}

\usepackage{graphicx}
\usepackage{newtxtext}
\usepackage{newtxmath}
\usepackage{natbib}
\usepackage{hyperref}
\usepackage{bm}

\hypersetup{
    colorlinks = true,
    urlcolor   = blue,
    citecolor  = black,
}

\newcommand{\RomanNumeralCaps}[1]
\linenumbers

\title{Dissipation-driven champion solitons in one-dimensional shallow-water waves}

\author{Ashleigh Simonis\aff{1}
  \corresp{\email{asimonis@umich.edu}},
  Sergey Nazarenko\aff{2},
  Jalal Shatah\aff{3}
 \and Yulin Pan\aff{1}\corresp{\email{yulinpan@umich.edu}}}

\affiliation{\aff{1}Department of Naval Architecture and Marine Engineering, University of Michigan, 2600 Draper Drive, Ann Arbor, Michigan 48109, USA
\aff{2}Institut de Physique de Nice, Université Côte d'Azur et CNRS, 17 rue Julien Lauprêtre, 06200 Nice, France
\aff{3}Courant Institute of Mathematical Sciences, New York University, 251 Mercer Street, New York, New York 10012, USA}

\begin{document}
\maketitle

\begin{abstract}
In this paper, we identify a new mechanism for rogue wave formation in a shallow-water setting. We study a bidirectional shallow-water wave field in the context of the Kaup-Boussinesq equation, and introduce a weak high-wavenumber dissipative perturbation that breaks the underlying integrability of the system. In this setting, dominant solitons grow through successive interactions with weaker, co-propagating solitons, leading to the formation of a ``champion soliton" in each direction of propagation. This behaviour is in contrast to the general intuition that dissipation damps coherent structures, and instead shows that weak dissipation can induce their intensification. Moreover, we find that weak dissipation alone is not sufficient for champion soliton formation; the presence of random waves plays a crucial role in the intensification process, catalysing the transfer of energy into dominant coherent structures. While champion solitons have previously been studied in non-integrable systems, these works primarily consider perturbations introduced through modifications of the nonlinear terms (e.g., higher-order Korteweg-de Vries and Schr\"{o}dinger-type models). In the present work, high-wavenumber dissipation provides a more physically natural perturbation, since such small-scale damping is a common feature in many systems.
\end{abstract}


\section{Introduction}

Solitary waves are a fundamental feature of nonlinear wave dynamics, emerging as coherent, strongly localised structures sustained by the interplay between nonlinearity and dispersion. Their study dates back to John Scott Russell's observation of a ``wave of translation" in a shallow-water canal \citep{russell1844}, and was later formalised through the Korteweg-de Vries equation, which admits solitary wave solutions \citep{kdv1895}. This phenomenon was revisited in the seminal work of \citet{zabusky1965}, where the term ``soliton" was introduced, and deeper insight into their nonlinear dynamics was provided. Since then, solitons have been a rich topic of interest within the nonlinear wave community, as they are relevant to a diverse range of physical settings, including water waves \cite[e.g.,][]{wu1998,guyenne2006}, nonlinear optics \cite[e.g.,][]{solli2007,kibler2010}, plasma physics \cite[e.g.,][]{Ichikawa1977,KUZNETSOV1986,Petviashvili2016}, Bose-Einstein condensates \cite[e.g.,][]{Strecker2002,Khaykovich2002}, and even biological systems such as nerve pulse propagation \cite[e.g.,][]{heimberg2005,Lautrup2011,heimburg2022}. Their role in modelling extreme ocean events, such as tsunamis and rogue waves \cite[e.g.,][]{kharif2003,clamond2006, dysthe2008,chabchoub2011,onorato2013}, is of particular practical importance. 

Interactions between solitons are a fundamental physical phenomenon in integrable systems. Such systems possess infinitely many conserved quantities and are exactly solvable via methods such as the inverse scattering transform (IST) \citep{ablowitz1985}. Representative examples include the KdV \cite[e.g.,][]{novikov1984,aktosun2004}, nonlinear Schr\"{o}dinger (NLS) \cite[e.g.,][]{karpman1981,Ablowitz2004,salem2009}, and sine-Gordon equations \cite[e.g.,][]{Hirota1972,karpman1981,dmitriev2014}. In such systems, solitons exhibit particle-like behaviour, meaning they can collide and recover their original shape and speed, with the interaction resulting only in a phase shift \citep{zabusky1965}. This elastic behaviour is a direct result of integrability, characterised by elastic scattering and isospectrality within the IST framework \cite[e.g.,][]{Eckhaus1981,ablowitz1985}.

When an integrable system is perturbed, elastic scattering is no longer preserved. In energy-conserving settings, such perturbations are typically introduced through the modification or addition of nonlinear terms \cite[e.g.,][]{kivshar1989,Besley2000,Omelyanov2016,Colleaux2025}, yielding non-integrable variants of the governing equations. Such perturbations render collisions inelastic or quasi-elastic, allowing irreversible energy transfer among solitons and interactions with surrounding random waves. This behaviour has been studied in the context of optical wave turbulence, where adding a nonlocal nonlinear correction to the NLS equation allows for non-integrable dynamics in which an initially turbulent state with coexisting random free waves and solitons evolves towards a single dominant soliton through successive inelastic collisions \cite[e.g.,][]{Bortolozzo2009,Laurie2012}. The emergence of such a dominant state, which we refer to here as a ``champion" soliton, was first described for non-integrable NLS equations with power-law or saturable nonlinearities \cite[e.g.,][]{zakharov1988,dyachenko1989,Jordan2000}, where it appears as the statistically favoured long-time coherent state of the system. The champion soliton phenomenon has been observed in the fourth-order generalised KdV equation (i.e., quartic instead of quadratic nonlinearity) \citep{flamarion2026}, providing a further example in a conservative non-integrable system.

The purpose of this work is to show that the champion soliton phenomenon can also exist in a 1-D bidirectional shallow-water setting. In particular, we model a bidirectional shallow-water random wave field via the completely integrable Kaup-Boussinesq (KB) model \cite[e.g.,][]{Kaup1975,Kupershmidt1985,El2001,Ivanov2009,Bhrawy2013,Gong2022}, perturbed by weak dissipation at high wavenumbers. Since high-wavenumber dissipation is present in many settings, it provides a more natural perturbation than the aforementioned studies that require modifying the nonlinearity. Our results show an initial relaxation stage, during which the energy of both random waves and solitons (identified via direct scattering transform) is dissipated. Afterwards, a champion soliton emerges in each direction of propagation, absorbing energy from smaller solitons during collisions. The champion solitons then propagate steadily with nearly constant velocity and amplitude before gradually decaying due to dissipative effects. The emergence of a champion soliton under a dissipative perturbation is noteworthy for two reasons: i) it is counterintuitive, since dissipative effects are generally expected to weaken solitons and damp oscillatory structures \citep{kivshar1989}, and ii) its occurrence in a system representative of a natural setting suggests a new mechanism for generating rogue waves in shallow water.

We finally remark that the behaviour observed is fundamentally different from dissipative solitons, which are stable localised structures that arise and are sustained by a balance of nonlinearity, dispersion, dissipation, and gain \cite[e.g.,][]{christov1995,Renninger2008,Purwins2010,Grelu2012}, as the absence of forcing in the system precludes such structures.

\section{The KB system \& numerical procedure}\label{num-pro}
We begin by considering the integrable Kaup-Boussinesq (KB) system in dimensionless form without dissipation \citep{Kaup1975}
\begin{equation}
    \eta_t + u_x + \alpha(\eta u)_x = -\frac{1}{3}\beta u_{xxx},
\label{eq:kb_n1}
\end{equation}
\begin{equation}
    u_t+\eta_x+\alpha uu_x=0,
\label{eq:kb_n2}
\end{equation}
where $\eta(x,t)$ is the surface elevation and $u(x,t)$ is the horizontal velocity. The parameters $\alpha=a/h$ and $\beta=(h/\lambda_p)^2$ are measures of the nonlinearity and dispersion, respectively, where $a$ represents the characteristic wave amplitude (e.g., half the significant wave height $H_s$), $h$ is the characteristic depth, and $\lambda_p$ is the peak wavelength. We take $\alpha=\beta=0.2$ throughout the paper.
In the absence of forcing and dissipation, the KB system conserves the total Hamiltonian \citep{Ali2017}
\begin{equation}
H =\int \left(\frac{\alpha^2}{2} \eta^2 + \frac{\alpha^2}{2} (1 + \alpha \eta) u^2 + \frac{\alpha^2 \beta}{3} u u_{xx} + \frac{\alpha^2 \beta}{6} u_x^2\right)dx.
\label{ham}
\end{equation}

A small-scale dissipative perturbation is introduced through a damping term on the right-hand side of \eqref{eq:kb_n1} and \eqref{eq:kb_n2}. In Fourier space, the system becomes
\begin{equation}
    \frac{\partial \hat{\eta}_k}{\partial t} =- i(k-\frac{1}{3}\beta k^3) \hat{u}_k + i\alpha \sum_{k_1+k_2=k} k \hat{\eta}_{k_1} \hat{u}_{k_2} -D_k\hat{\eta}_k,
\label{eq:kb_F1}
\end{equation}
\begin{equation}
    \frac{\partial \hat{u}_k}{\partial t} =- ik \hat{\eta}_k - \frac{i}{2} \alpha \sum_{k_1+k_2=k} k  \hat{u}_{k_1} \hat{u}_{k_2} -D_k\hat{u}_k,
\label{eq:kb_F2}
\end{equation}
with
\begin{equation}
    D_k=\frac{1}{\Delta t}\left|\frac{k}{\beta_1k_p}\right|^{\beta_2},
\end{equation}
where $\beta_1=4$ and $\beta_2=20$, $\Delta t$ denotes the time step, and $k_p$ denotes the peak wavenumber. The coefficient $D_k$ is defined in such a way to be consistent with the filter applied in \citep{Xiao2013}, where the Fourier coefficients are multiplied by
\begin{equation}
    \Lambda(k | k_p, \beta_1,\beta_2)=\exp\bigg(-\bigg\lvert \frac{k}{\beta_1k_p}\bigg\rvert^{\beta_2}\bigg),
\label{orig_dissip}
\end{equation}
providing weak smoothing at high wavenumbers, while leaving large scales unaffected.

We simulate the KB system \eqref{eq:kb_F1} and \eqref{eq:kb_F2} using a pseudospectral method paired with a fourth-order Runge-Kutta time-marching scheme with an integration factor (IF-RK4) formulation \citep[e.g.,][]{Pan2020,simonis2026}. Numerical experiments of the bidirectional wave field are performed on a periodic computational domain of size $L=1024\pi$ with $N=4096$ free wave modes (before de-aliasing). We initialise simulations using a Gaussian spectrum $S(k)$ \citep[e.g.,][]{Pelinovsky2006,Flamarion2024} 
\begin{equation}
    S(k)=Q\exp\left(-\frac{(k-k_p)^2}{2K^2}\right),
\end{equation}
where the peak wavenumber is set to $k_p=1$, while $K=0.1$ controls the spectral bandwidth. The parameter $Q$ is chosen so that the resulting data remain of $O(1)$, as measured by $||\eta||_\infty$ and $||u||_\infty$. The amplitude of each Fourier mode $\hat{\eta}_k$ is drawn from the spectrum $S(k)$. To generate a bidirectional field, the energy is evenly divided by splitting $\hat{\eta}_k$ between left- and right-propagating components. Each component is assigned an independent random phase drawn uniformly from $(0,2\pi]$. The corresponding velocity component $\hat{u}_k$ is then obtained from $\hat{\eta}_k$ for both propagation directions using the linear wave theory relation. We simulate the system for a long evolution time, i.e., $t=O(10^6T_p)$, where $T_p$ is the peak period associated with $k_p$, to capture sufficient wave-wave interactions and successive solitary wave collisions under the perturbation.

We now formally introduce the solitary waves supported by the system, defining solitons through the direct scattering transform (DST), also known as the nonlinear Fourier transform. Although the full IST framework involves both a spectral (linear eigenvalue) problem and a time-evolution equation, the present study focuses solely on the former. At each instant of time, the spectral problem for the KB system is given by \citep{Kaup1975}
\begin{equation}
\Psi_{xx}+\left[E(\zeta)^2+ik(\zeta)q(x)+r(x)\right]\Psi=0,
\label{spec-zeta}
\end{equation} 
where $k(\zeta)=\tfrac{1}{4}\left(\zeta-\frac{B}{\zeta}\right)$ and $E(\zeta)=\tfrac{1}{4}\left(\zeta+\frac{B}{\zeta}\right)$ with $B=3/\beta$, and $\Psi$ is the eigenfunction. The potentials $q(x)$ and $r(x)$ are defined as
\[
q(x)=\frac{\sqrt3}{2}\alpha\beta^{-1/2}u(x), \quad r(x)=\frac{3}{4}\alpha\beta^{-1}\left(\eta(x)-\frac{1}{4}\alpha u(x)^2\right),
\]
which are assumed to decay to zero at $|x|\rightarrow \infty$. Since the transformation $\zeta\mapsto-B/\zeta$ leaves $k(\zeta)$ and $E(\zeta)^2$ invariant while mapping $E(\zeta)\mapsto-E(\zeta)$, each $\zeta$ solution of \eqref{spec-zeta} has a ``mirror point". To avoid double counting, we restrict our analysis to the upper half $E$-plane, i.e., $\mathrm{Im}[E(\zeta)]>0$. By solving \eqref{spec-zeta}, we are able to obtain a discrete spectrum of eigenvalues $\zeta_j$. Each bound state in the system corresponds uniquely to a discrete eigenvalue of the DST problem. We note that for $\beta>0$, bound states correspond either to purely imaginary eigenvalues ($\zeta=i\eta$, $\eta\in\mathbb{R}$), or to complex-conjugate pairs satisfying $\zeta_2=-\zeta_1^*$, as in the case of breathers in the sine-Gordon equation \citep{Kaup1975}. 

The one-soliton solution of the KB system with $\beta>0$ is given by \citep{Kaup1975,guyenne2006}
\begin{equation}
u(x,t)=\frac{2(c_s^2-1)}{\alpha\ \mathrm{cosh}(\sqrt{\frac{3}{\beta}(c_s^2-1)}(x-c_st))+c_s},
\label{u-sol}
\end{equation}
\begin{equation}
\eta(x,t)=c_su(x,t)-\frac{\alpha}{2}u(x,t)^2.
\label{eta-sol}
\end{equation}
This is a single-parameter solution, parameterised by the soliton velocity $c_s$. Here, $c_s=\omega_s/\kappa_s$, where
\begin{equation}
\omega_s=-\frac{\sqrt{\beta}}{8\sqrt{3}}
\left(\zeta^2-\frac{9}{\beta^2\zeta^2}\right),
\qquad
\kappa_s=-\frac{i}{4}\left(\zeta+\frac{3}{\beta\zeta}\right).
\label{sol_param}
\end{equation}
Evidently, the key characteristics of solitons (e.g., width, amplitude, velocity) are determined by the corresponding value of $\zeta$. We note that the standard DST procedure assumes the problem is defined on $\mathbb{R}$, whereas our numerical solutions for $u(x)$ and $\eta(x)$ are periodic. For this reason, the DST may not be exact; however, it remains an effective tool for detecting bound states in the solution (see \cite{Colleaux2025} as an example). From \eqref{u-sol} and \eqref{eta-sol}, we can see that for a soliton to be physical and realisable, it requires $|c_s|>1$. In the $\zeta$-plane, the continuous spectrum corresponds to $\textnormal{Im}[E(\zeta)]=0$, i.e., $\zeta\in\mathbb{R}$ or $|\zeta|=\sqrt{3/\beta}$ (see caption of Fig. \ref{fig:disc_spec_cons} for details). As discrete eigenvalues approach the continuous spectrum, the corresponding velocity approaches 1. Thus, we only retain discrete eigenvalues which satisfy $|c_s|>1$ and are separated from the continuous spectrum curve by at least a 1\% buffer, $\big||\zeta|-\sqrt{3/\beta}\big|>0.01\sqrt{3/\beta}$. Given our finite, yet sufficiently large box size, these retained solitons align with the definition of ``true" bound states where the tails decay to zero as $|x|\rightarrow\infty$. Full details of the DST procedure can be found in \citep{simonis2026}, and a validation of soliton solutions is provided in Appendix \ref{appA}.

\section{Results}
\subsection{The non-dissipated KB system}\label{cons-kb}

\begin{figure}
\centerline{\includegraphics[]{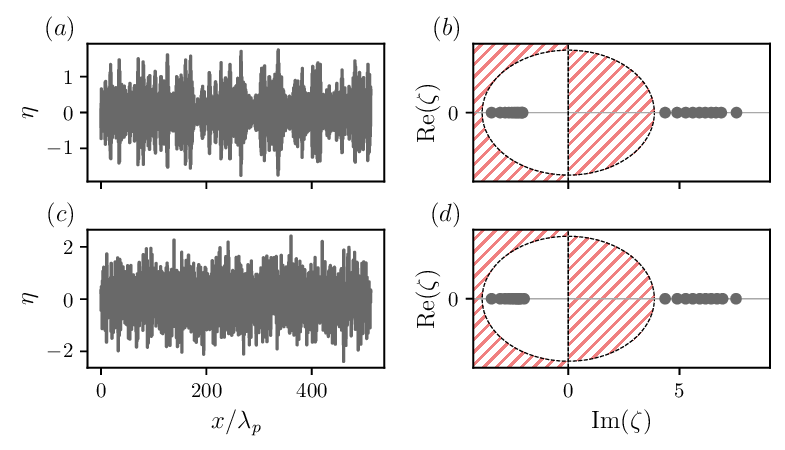}}
\caption{Surface elevation $\eta$ and corresponding discrete eigenvalue spectrum in the $\zeta$-plane obtained via DST at (top row) $t=0$ and (bottom row) $t\approx2\times10^6T_p$ for the conservative KB system. The lower half $E$-plane (i.e., where $\textnormal{Im}[E(\zeta)]<0$) is marked by the red hatched region and the continuous spectrum region (i.e., where $\textnormal{Im}[E(\zeta)]=0$) is marked by the dashed black line.}
\label{fig:disc_spec_cons}
\end{figure}

We begin by presenting results for the freely evolving KB system \eqref{eq:kb_n1} and \eqref{eq:kb_n2}, i.e., in the absence of dissipation, as a baseline for the integrable behaviour of the system. At the beginning of the simulation, the DST identifies solitons embedded in the random wave background, with the initial field shown in Fig. \ref{fig:disc_spec_cons}$(a)$ and the corresponding discrete spectrum shown in figure \ref{fig:disc_spec_cons}$(b)$. Figure \ref{fig:disc_spec_cons}$(c,d)$ shows that at a later time, the field remains random and the discrete spectrum remains unchanged, with the latter reflecting the isospectrality of the system. Here, eigenvalues with $\mathrm{Im}(\zeta)<0$ correspond to right-propagating solitons, while those with $\mathrm{Im}(\zeta)>0$ correspond to left-propagating solitons.

\subsection{Champion solitons in the dissipated KB system}\label{sol-evo}
\begin{figure}
\centerline{\includegraphics[]{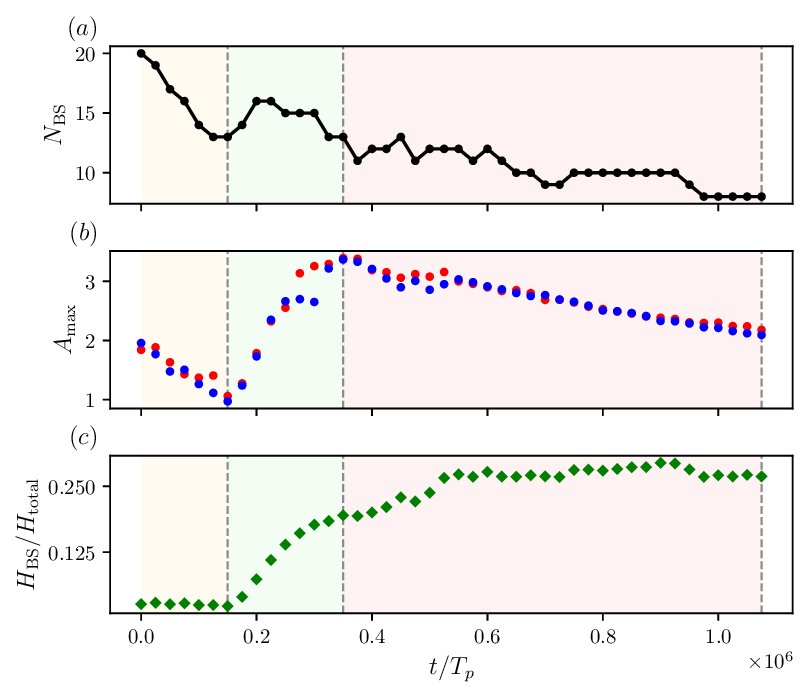}}
\caption{Evolution of bound-state properties of the system obtained via DST. $(a)$ Number of bound states $N_\mathrm{BS}$ present in the system; $(b)$ maximum amplitude of the left-propagating (red points) and right-propagating (blue points) solitons; $(c)$ ratio of soliton energy to the total system energy. The relaxation, intensification, and quasi-steady stages, including subsequent decay, are marked by orange, green, and red shading, respectively.
}
\label{fig:bs_evo}
\end{figure}

\begin{figure}
\centerline{\includegraphics[]{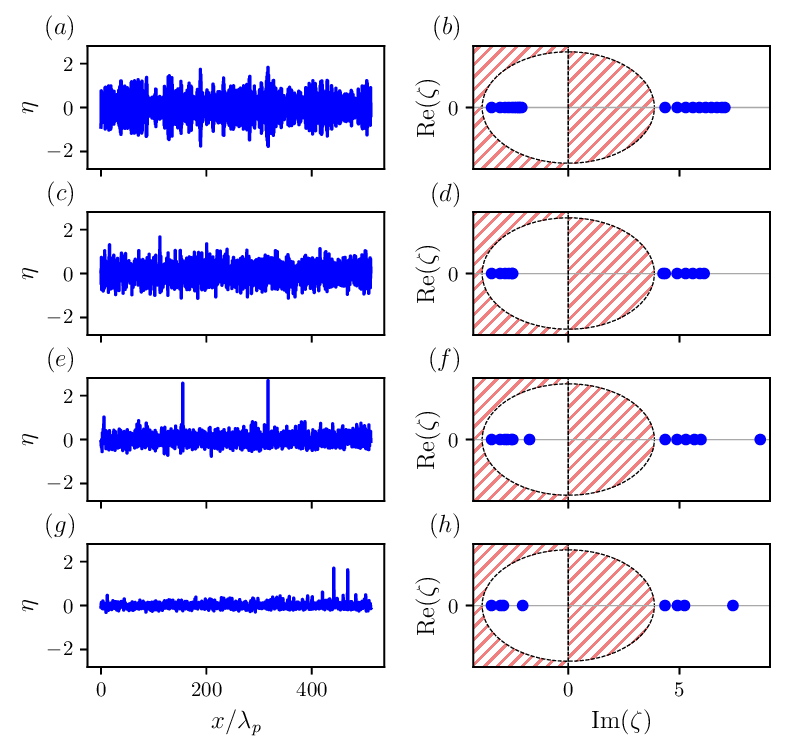}}
\caption{Evolution of the surface elevation $\eta$ (left column) and corresponding discrete eigenvalue spectrum in the $\zeta$-plane obtained via DST (right column). The four rows present snapshots at $t=0$, $t=1.5\times10^5T_p$, $t=3.5\times10^5T_p$, and $t=1.075\times10^6T_p$.}
\label{fig:disc_spec_relax}
\end{figure}

Next, we examine how the system behaves under the dissipative perturbation. Figure \ref{fig:bs_evo} illustrates the long-time behaviour of the system by showing the number and peak amplitude of the bound states obtained via DST, as well as their fraction of the total system energy computed using \eqref{ham}. The evolution and emergence of the champion solitons appear to follow three distinct stages, based on the soliton peak amplitudes shown in Fig. \ref{fig:bs_evo}$(b)$. We refer to these as the relaxation, intensification, and quasi-steady stages, after which the champion solitons ultimately succumb to dissipative effects. Each of these stages is further illustrated by Fig. \ref{fig:disc_spec_relax} which presents the field and corresponding discrete spectrum at four distinct times marking the thresholds of each stage.

\textbf{Relaxation stage -- Fig. \ref{fig:disc_spec_relax}$(a)$--$(d)$:}
The relaxation stage is characterised by an initial decrease in the number of bound states and peak bound-state amplitudes, as shown in figure \ref{fig:bs_evo}$(a,b)$. During this stage, both the random wave field and solitons are dissipated. The nearly constant energy fraction in Fig. \ref{fig:bs_evo}$(c)$ suggests that the solitons and wave field decay at the same rate. Figure \ref{fig:disc_spec_relax}$(a,c)$ shows snapshots of the wave field at the beginning and end of the relaxation stage, illustrating the decrease in field amplitude. At the same time, Fig. \ref{fig:disc_spec_relax}$(b,d)$ shows the corresponding discrete spectra, where the decrease in the number of discrete eigenvalues and their shift towards the continuous-spectrum threshold indicate weaker soliton amplitudes.

\textbf{Intensification stage -- Fig. \ref{fig:disc_spec_relax}$(c)$--$(f)$:} The intensification stage starts after the solitons have dissipated to a low-intensity state, with a reduction in number and peak amplitude, as shown in Fig. \ref{fig:bs_evo}$(a,b)$. In Fig. \ref{fig:disc_spec_relax}$(c)$, the solitons remain embedded in the random wave field, while the corresponding discrete spectrum in Fig. \ref{fig:disc_spec_relax}$(d)$ reveals that the eigenvalues are closely spaced at this time, reflecting a configuration of co-propagating solitons with comparable amplitudes. Figure \ref{fig:bs_evo}$(b)$ demonstrates that at later times in this stage, the peak amplitudes continue to grow, while the number of bound states gradually decreases. Meanwhile, we see in Fig. \ref{fig:bs_evo}$(c)$ that the soliton energy fraction steadily increases over this interval. The stage ends when the strongest bound states reach their maximum intensity in Fig. \ref{fig:bs_evo}$(b)$, marking the point at which dissipative losses hinder further amplification. Figure \ref{fig:disc_spec_relax}$(e)$ shows that the stage culminates with the emergence of a ``champion soliton" in each propagation direction, which outpaces the subdominant solitons and becomes the strongest coherent structure in the system. At this time, $\eta_{\max}/H_s\approx3$, where $H_s=4\sigma_{\eta}$ is the significant wave height, classifying the champion soliton as a rogue wave under the crest-based criterion of $\eta_{\max}/H_s>1.25$ \cite[e.g.,][]{Cattrell2018}. This behaviour is reflected in the discrete spectrum in Fig. \ref{fig:disc_spec_relax}$(f)$, where we see one discrete eigenvalue in each propagation direction that exhibits pronounced intensification and separates from the rest.

\textbf{Quasi-steady stage -- Fig. \ref{fig:disc_spec_relax}$(e)$--$(h)$:}
Following the intensification phase, the champion solitons experience a competition between energy loss due to dissipative effects and energy gain from interactions with subdominant solitons. Figure \ref{fig:bs_evo}$(b)$ shows that at early times in this stage, there is an interval of $O(10^5T_p)$ where the effects are nearly balanced, and the champion solitons propagate with nearly constant amplitudes. Meanwhile, Fig. \ref{fig:bs_evo}$(c)$ shows that the soliton energy fraction continues to increase, indicating that the soliton component becomes increasingly dominant, despite the absence of further champion-soliton intensification. At later times, the energy fraction approaches a plateau as the champion-soliton intensities begin to decrease, as evident in Fig. \ref{fig:bs_evo}$(b,c)$. This indicates that further growth and preservation of the champion solitons are limited. The decrease in champion-soliton intensity and random-wave background over this stage is shown in Fig. \ref{fig:disc_spec_relax}$(e,g)$. The corresponding discrete spectra in Fig. \ref{fig:disc_spec_relax}$(f,h)$ show both reductions in the intensities of the champion-soliton eigenvalues and the number of discrete subdominant eigenvalues.

\begin{figure}
\centerline{\includegraphics[]{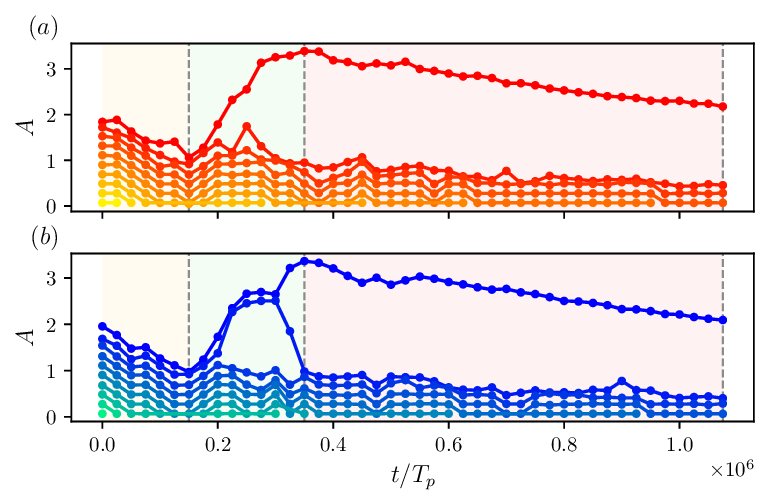}}
\caption{Evolution of the amplitudes of the bound states present in the system. $(a)$ Left-propagating solitons; $(b)$ right-propagating solitons. Each curve represents an individual bound state, with darker colours indicating stronger solitons. The relaxation, intensification, and quasi-steady stages, including subsequent decay, are marked by orange, green, and red shading, respectively.}
\label{fig:amp_evo}
\end{figure}

The full process is also shown in Fig. \ref{fig:amp_evo}, which tracks the evolution of each soliton in the system. We see that at the end of the relaxation stage, the soliton amplitudes are more comparable than in the initial condition. At the onset of intensification, some subdominant solitons initially grow; however, the subsequent growth disproportionately favours one soliton in each direction of propagation, which draws energy from the others. At the end of the intensification stage, the champion soliton in each direction is well separated in amplitude from the remaining subdominant solitons. We note that while the number of subdominant solitons decreases, a few smaller solitons persist throughout the quasi-steady stage. This likely reflects two effects: i) interactions with the champion soliton occur over very short timescales due to the large velocity difference, making energy transfer inefficient, and ii) the subdominant bound states have amplitudes and velocities comparable to those of the surrounding random waves, which may allow energy exchange with the random-wave background and thus support their long-time persistence.

An animation showing the evolution of the wave field and corresponding discrete spectra is provided in the Supplementary Material. We remark that the same qualitative long-time behaviour is observed across different realisations with distinct initial random phases. We also note that slightly increasing the dissipation strength appears to accelerate the formation of champion solitons, as shown in Appendix \ref{appC}. If the dissipation is increased further or shifted to larger scales, the solitons may be rapidly damped out. The present configuration remains weakly perturbed, maintaining the balance required for long-lived soliton dynamics.

\subsection{Mechanism for soliton intensification}
To assess the role of random waves in the soliton intensification process, we perform two experiments: one with random waves and one without. In the first experiment, we initialise an ensemble of counter-propagating solitons with varied amplitudes constructed using \eqref{u-sol} and \eqref{eta-sol} and allow it to evolve according to \eqref{eq:kb_F1} and \eqref{eq:kb_F2}. In the second experiment, we simulate the same system and soliton ensemble, but with an added bidirectional random-wave component.

\begin{figure}
\centerline{\includegraphics[]{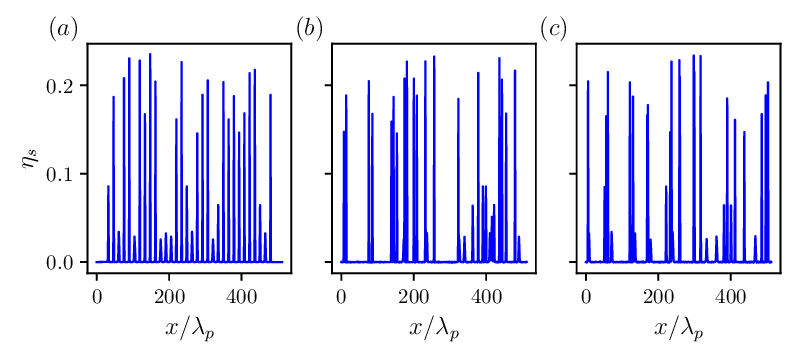}}
\caption {Evolution of the soliton-only component $\eta_s$ under weak small-scale dissipation. Panels $(a)$--$(c)$ show the 32-soliton ensemble at $t=0$, $t=3.75\times10^5T_p$, and $t=7.5\times10^5T_p$, respectively.}
\label{fig:sol_exp}
\end{figure}

Figure \ref{fig:sol_exp} shows the evolution in the first experiment. The snapshots demonstrate that no champion soliton emerges during the process; moreover, the ensemble propagates essentially unchanged over several thousand periods. This behaviour indicates that the collisions are nearly elastic.

\begin{figure}
\centerline{\includegraphics[]{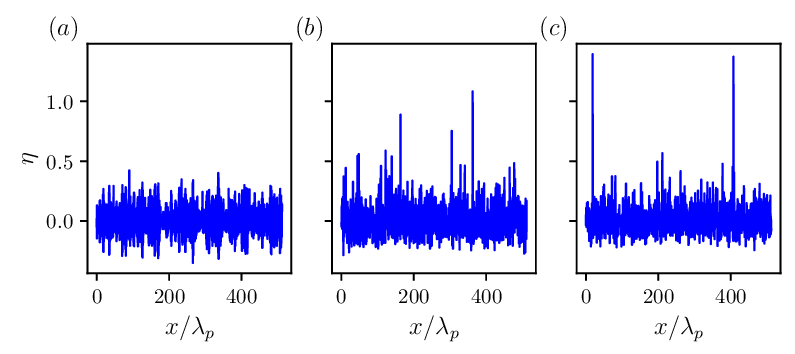}}
\caption {Evolution of the surface elevation $\eta=\eta_s+\eta_{rw}$, under weak small-scale dissipation, where $\eta_s$ is the 32-soliton ensemble in Fig. \ref{fig:sol_exp} and $\eta_{rw}$ is a bidirectional random wave component. Panels $(a)$--$(c)$ show the field at $t=0$, $t=3.75\times10^5T_p$, and $t=7.5\times10^5T_p$, respectively.}
\label{fig:sol_rw_exp}
\end{figure}

Figure \ref{fig:sol_rw_exp} shows that the behaviour in the second experiment is notably different from the first, and consistent with that observed in \S \ref{sol-evo}. In particular, the evolution results in the emergence and long-lived propagation of champion solitons. This case provides evidence that dissipation alone is not sufficient to induce the amplification required for the formation of champion solitons. While a dissipative perturbation is evidently necessary, as shown in \S \ref{cons-kb}, the presence of the random-wave component appears to serve as a catalyst for this process. Since the weak dissipative perturbation targets high wavenumbers, its effect directly depends on the small-scale spectral content. In the KB system, interactions among random waves can rapidly populate small scales via quasi-resonant interactions \citep{simonis2026}. Since the dissipation considered here is weak, the initial growth of such scales is not suppressed. The resulting high-wavenumber content then provides the spectral component required for the dissipative perturbation to act effectively.

\begin{figure}
\centerline{\includegraphics[]{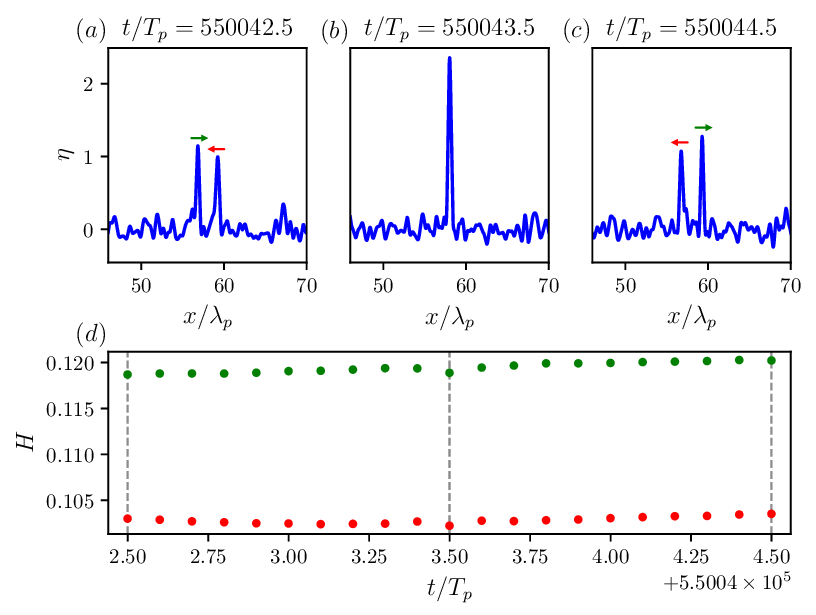}}
\caption{Example of a counter-propagating soliton collision in the second experiment. Panels $(a)$--$(c)$ show the solitons before, during, and after the collision, respectively, with arrows indicating the direction of propagation. Panel $(d)$ tracks the energy of the right-propagating soliton (green points) and the left-propagating soliton (red points) over the collision interval, with dashed vertical lines marking the times shown in panels $(a)$--$(c)$.}
\label{fig:countprop}
\end{figure}

Having established that the presence of a random-wave component is essential for the development of champion solitons, we next examine which type of soliton interaction contributes most significantly to the energy transfer. To this end, we measure the energy of each soliton during two different interaction types observed in the second experiment, namely collisions between counter-propagating solitons and overtaking interactions between co-propagating solitons. Figure \ref{fig:countprop}$(a$--$c)$ shows a counter-propagating collision between two dominant solitons. During the interaction, the solitons briefly merge into a single-crest structure before separating and continuing on their respective trajectories. This interaction is short-lived, i.e., $O(1T_p)$, and the solitons emerge from the collision with their profiles virtually unchanged. At each time in Fig. \ref{fig:countprop}$(d)$, the energy of each colliding soliton is computed using \eqref{ham}, with the soliton profiles reconstructed from their associated discrete eigenvalues. No significant energy transfer is observed, likely because the timescale of the interaction is too short.

\begin{figure}
\centerline{\includegraphics[]{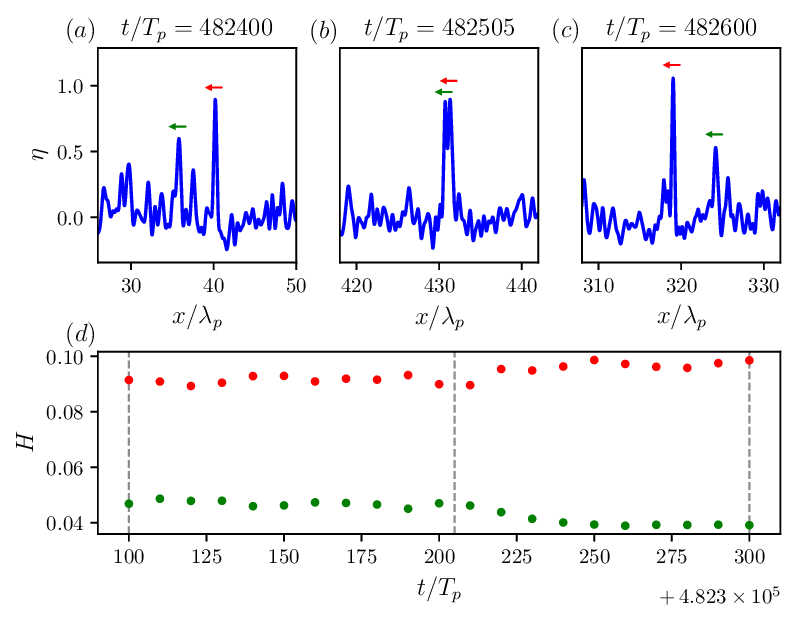}}
\caption{Example of a co-propagating soliton collision in the second experiment. Panels $(a)$--$(c)$ show the left-propagating solitons before, during, and after the collision, respectively, with the red and green arrows indicating the dominant and subdominant solitons. Panel $(d)$ tracks the energy of the dominant soliton (red points) and the subdominant soliton (green points) over the collision interval, with dashed vertical lines marking the times shown in panels $(a)$--$(c)$.}
\label{fig:coprop}
\end{figure}

Next, we look at a co-propagating collision in Fig. \ref{fig:coprop}$(a$--$c)$. Here, the dominant left-propagating soliton ``catches" the slower, subdominant soliton. The pair propagates together for an extended period, i.e., $>O(10T_p)$. During this period, while the two solitons are coupled, they retain two well-defined and separated crests. For the amplitude ratio of the interacting pair considered, the Lax-type classification predicts a two-crest collision structure, consistent with the behaviour observed here \cite[e.g.,][]{lax1968,zhang2003}. Eventually, the dominant soliton detaches, and two well-separated soliton profiles re-emerge. Figure \ref{fig:coprop}$(d)$ reveals that the dominant soliton gains energy during the interaction, while the subdominant soliton loses it. The prolonged joint propagation likely gives the dominant soliton time to draw energy from the subdominant one, allowing this interaction type to be most effective at transferring energy. This supports the notion that energy transfer is most efficient when many co-propagating solitons with similar amplitudes are present, since closer amplitudes and velocities lead to longer joint propagation times, as shown in Appendix \ref{appB}. These prolonged interactions may provide a mechanism for triggering the intensification stage in \S \ref{sol-evo}. The observation of the phenomenon in the unidirectional KdV case in Appendix \ref{appD} supports this interpretation.

\section{Discussion \& Conclusion}
We study the emergence and long-time properties of dissipation-driven champion solitons in the KB system. Starting from a random bidirectional wave field, introducing a dissipative perturbation breaks the complete integrability of the system, resulting in a pronounced intensification of certain solitons over time. This behaviour is in contrast to the non-dissipated KB system, where isospectrality ensures elastic scattering, preserving the soliton ensemble without preferential growth. Our results demonstrate that dissipation acting on a soliton-only ensemble is insufficient to give rise to champion solitons, and that the presence of a random-wave component plays a key role.

The influence of the random-wave component may be twofold. First, it provides small-scale spectral content for the high-wavenumber weak dissipation to act on. This introduces the effective perturbation, allowing for the departure from isospectrality and associated elastic scattering. Second, in \S \ref{sol-evo} we note that although the two champion solitons are the most prominent coherent structures in the field, a few very weakly bound states persist. These low-amplitude, low-velocity bound states remain embedded in the random-wave background. Their persistence suggests that the random waves may act as an energy reservoir, sustaining smaller solitons and offsetting the energy deficits they incur during interactions with larger solitons.

The findings in this work are particularly relevant to physical shallow-water settings where random waves, coherent structures, and dissipation coexist. Dissipation is typically expected to suppress extreme wave events by damping energy in the system. Our results reveal a counterintuitive mechanism: in the presence of random waves, dissipation drives the emergence of dominant solitons. By targeting small-scale components of the wave field, dissipation enables energy transfer that amplifies dominant solitons rather than suppressing them. This unexpected role of dissipation is of particular interest as it identifies a new mechanism for rogue wave generation in shallow-water environments.

\backsection[Supplementary data]{\label{SupMat}Supplementary animation is available at ...}

\backsection[Acknowledgements]{The authors would like to thank Prof. Peter Miller for the constructive discussion and valuable insight on this topic.}

\backsection[Funding]{This research was supported by the Simons Foundation (Award ID \#651459). The numerical simulations were performed on the Great Lakes HPC Cluster provided by Advanced Research Computing (ARC) at the University of Michigan, Ann Arbor.}

\backsection[Declaration of interests]{The authors report no conflict of interest.}

\backsection[Author ORCIDs]{A. Simonis, https://orcid.org/0009-0000-3876-8160; S. Nazarenko, https://orcid.org/0000-0002-8614-4907; J. Shatah, https://orcid.org/0000-0003-3518-8069; Y. Pan, https://orcid.org/0000-0002-7504-8645}

\appendix

\section{Reconstruction of solitons from discrete eigenvalues}\label{appA}

\begin{figure}
\centerline{\includegraphics[]{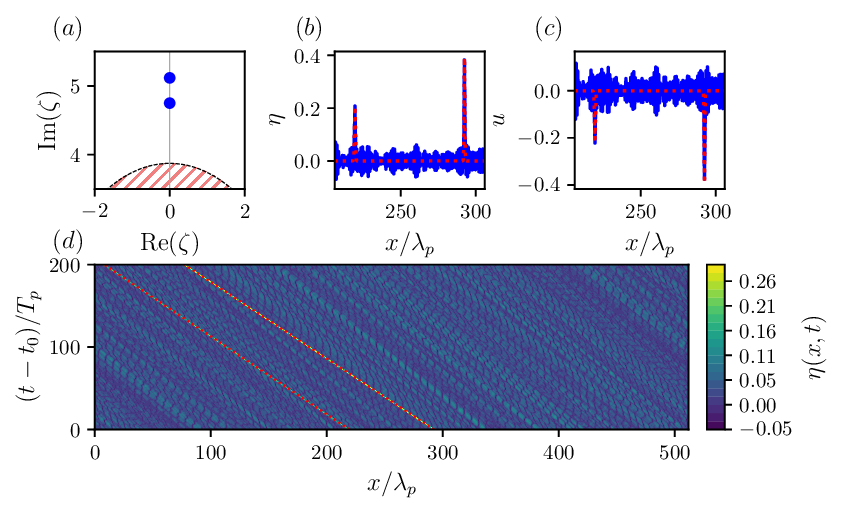}}
\caption{$(a)$ Discrete spectrum with two eigenvalues corresponding to two left-propagating solitons. $(b)$ Surface elevation from the numerical simulation (solid blue) with the soliton profiles reconstructed using \eqref{eta-sol} (dotted red). $(c)$ Horizontal velocity from the numerical simulation (solid blue) with the soliton profiles reconstructed using \eqref{u-sol} (dotted red). $(d)$ Space-time surface elevation plot with soliton trajectories indicated by dotted red lines computed from soliton velocities $c_1$ and $c_2$, respectively.}
\label{fig:validation}
\end{figure}

In this appendix, we validate the DST by ensuring that the solitons present in the system are in correspondence with the set of discrete eigenvalues. Full details of the procedure for solving the eigenvalue problem can be found in \cite{simonis2026}. Figure \ref{fig:validation}$(a)$ shows an example discrete spectrum with two eigenvalues $\zeta_1$ and $\zeta_2$, each corresponding to a left-propagating soliton. Using \eqref{sol_param} with $\beta=0.2$, we obtain soliton velocities $c_1=1.02$ and $c_2=1.04$, respectively. 
With these velocities, we can then construct the soliton profiles using \eqref{u-sol} for horizontal velocity and \eqref{eta-sol} for surface elevation. Figure \ref{fig:validation}$(b,c)$ shows good agreement between the solitary wave structures identified in our numerical data and the reconstructed profiles. Additionally, the soliton trajectories in Fig. \ref{fig:validation}$(d)$ are confirmed by overlaying the predicted path $x_i(t)=x_i(t_0)-c_i(t-t_0)$ for $i=1,2$.

\section{Influence of dissipation strength}\label{appC}
\begin{figure}
\centerline{\includegraphics[]{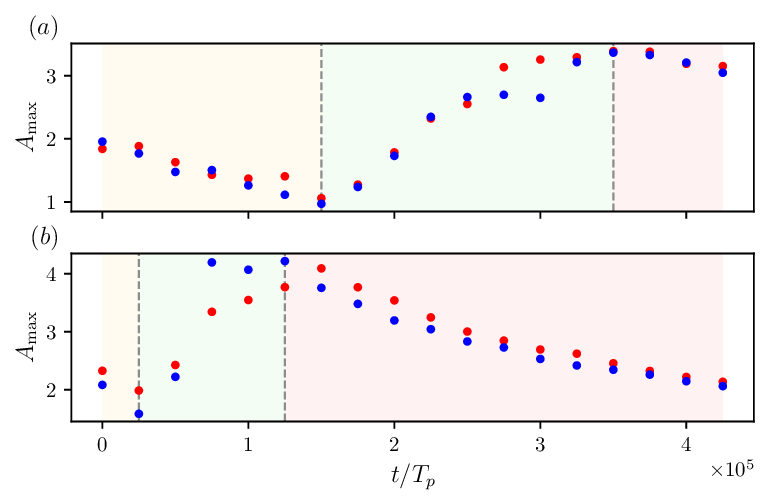}}
\caption{Evolution of the maximum amplitude of the left-propagating soliton (red points) and right-propagating soliton (blue points). Panel $(a)$ corresponds to the dissipation settings used in the main paper, and panel $(b)$ corresponds to the stronger dissipation setting. The relaxation, intensification, and part of the quasi-steady stage are marked by orange, green, and red shading, respectively.}
\label{fig:strong-diss}
\end{figure}

This appendix examines the influence of dissipation strength on the system's dynamics. We simulate \eqref{eq:kb_F1} and \eqref{eq:kb_F2} using the same numerical setup as in the main paper, but with a slightly stronger dissipative perturbation. In this case, the dissipative coefficient $D_k$ is modified by setting $\beta_1=3$ and $\beta_2=30$. Figure \ref{fig:strong-diss} demonstrates that the system exhibits the same qualitative behaviour as in the main paper; however, the onset of intensification and the emergence of champion solitons appear to be accelerated.

\section{Dependence of interaction time on amplitude ratio}\label{appB}
\begin{figure}
\centerline{\includegraphics[]{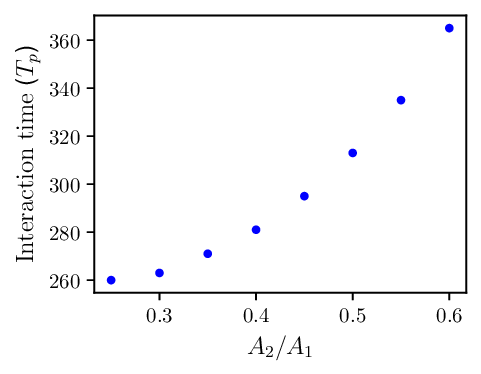}}
\caption{Interaction time as a function of amplitude ratio $A_2/A_1$ for solitons with $A_1>A_2$.}
\label{fig:amp_rat}
\end{figure}

In figure \ref{fig:amp_rat}, we provide an example illustrating how the interaction time depends on the amplitude ratio for collisions of two co-propagating solitons. The interaction time is defined as the duration between the initial overlap of the soliton tails and their subsequent separation. We observe that the interaction time increases monotonically with the amplitude ratio.

\section{Unidirectional KdV simulation}\label{appD}
By eliminating one direction of wave propagation, the KB system \eqref{eq:kb_n1} and \eqref{eq:kb_n2} can be used to derive a KdV equation, as described in \cite{Karczewska2018} (see Ch. 3). The resulting KdV equation is
\begin{equation}
    \eta_t+\eta_x+\frac{3}{2}\alpha\eta\eta_x+\frac{1}{6}\beta\eta_{xxx}=0.
    \label{kdv-app}
\end{equation}

\begin{figure}
\centerline{\includegraphics[]{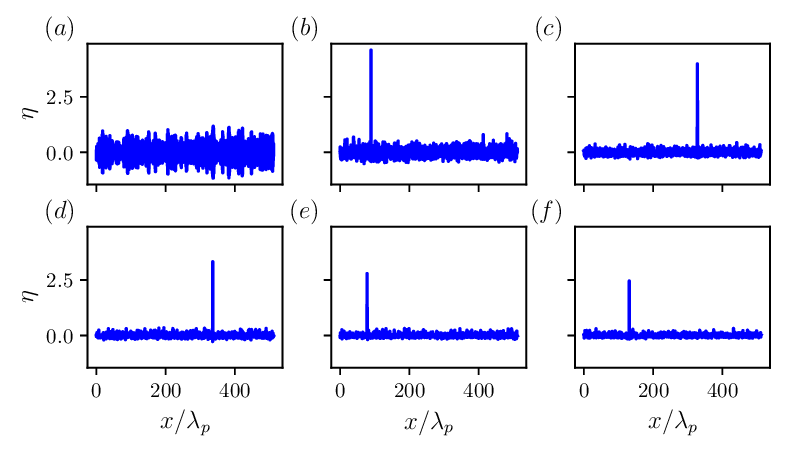}}
\caption{Evolution of the surface elevation $\eta$ of the dissipated KdV equation. Panels $(a)$--$(f)$ show snapshots at $t=0$, $t=2\times10^5T_p$, $t=4\times10^5T_p$, $t=6\times10^5T_p$, $t=8\times10^5T_p$, and $t=10^6T_p$, respectively.}
\label{fig:uni}
\end{figure}

We simulate \eqref{kdv-app} using a pseudospectral method combined with an IF-RK4 time-marching scheme, consistent with the main paper. The simulation uses the same configuration detailed in \S\ref{num-pro}, with the only difference being that the initial condition now only includes right-propagating waves. Figure \ref{fig:uni} demonstrates that starting from a unidirectional random wave field and allowing it to evolve under weak small-scale dissipation also leads to champion-soliton behaviour. In this case, however, only one soliton emerges as there is only one direction of propagation. We also note that the emergence of the champion soliton appears to occur more rapidly than in the bidirectional case, likely because only co-propagating collisions occur, allowing energy to be concentrated more efficiently into the dominant soliton.
\FloatBarrier
\bibliographystyle{jfm}
\bibliography{jfm}

\end{document}